\documentclass[trackchanges,twocolumn,resetfootnote]{aastex701}
\usepackage{graphics,graphicx}
\hypersetup{urlcolor=blue}
\usepackage{natbib}
\usepackage{textcomp,gensymb}
\usepackage{xfrac}
\usepackage{xcolor}
\usepackage{multirow}

\newcommand{\vsini}{$v\sin{i_*}$}

\newcommand{\teff}{\ensuremath{T_{\text{eff}}}}
\newcommand\kms{km~s$^{-1}$}
\newcommand{\ms}{m~s$^{-1}$}
\newcommand{\msd}{m~s$^{-1}$~d$^{-1}$}

\newcommand{\tess}{\textit{TESS}}

\newcommand{\maroonx}{\textit{MAROON-X}}
\newcommand{\muscat}{\textit{MuSCAT3}}
\newcommand{\lco}{\textit{LCOGT}}

\newcommand{\planetname}{IRAS 04125+2902\,b}
\newcommand{\starname}{IRAS 04125+2902}
\newcommand{\altplanetname}{TIDYE-1\,b}

\newcommand{\eso}{European Southern Observatory, Karl-Schwarzschildstraße 2, D-85748 Garching bei München, Germany}

\pdfoutput=1

\shorttitle{TIDYE-1b is (probably) aligned} 
\shortauthors{Barber et al.}

\begin{document}

\title{Stellar Obliquities of Young Systems, Atmospheres Undergoing Contraction and Escape (SOYSAUCE): a likely aligned orbit for the 3\,Myr planet TIDYE-1\,b}

\correspondingauthor{Madyson G. Barber}

\author[0000-0002-8399-472X]{Madyson G. Barber}
\altaffiliation{NSF Graduate Research Fellow}
\affiliation{Department of Physics and Astronomy, The University of North Carolina at Chapel Hill, Chapel Hill, NC 27599, USA} 
\email{madysonb@live.unc.edu}  

\author[0000-0003-3654-1602]{Andrew W. Mann}
\affiliation{Department of Physics and Astronomy, The University of North Carolina at Chapel Hill, Chapel Hill, NC 27599, USA}
\email{awmann@unc.edu}  

\author[0000-0002-5099-8185]{Marshall C. Johnson}
\affiliation{Department of Astronomy, The Ohio State University, 4055 McPherson Laboratory, 140 West 18$^{\mathrm{th}}$ Ave., Columbus, OH 43210 USA}
\email{johnson.7240@osu.edu}

\author[0000-0003-1368-6593]{Mayuko Mori}
\affiliation{Astrobiology Center, 2-21-1 Osawa, Mitaka, Tokyo 181-8588, Japan}
\affiliation{National Astronomical Observatory of Japan, 2-21-1 Osawa, Mitaka, Tokyo 181-8588, Japan}
\email{mayukomori.519@gmail.com} 

\author[0000-0002-4881-3620]{John Livingston}
\affiliation{Astrobiology Center, 2-21-1 Osawa, Mitaka, Tokyo 181-8588, Japan}
\affiliation{National Astronomical Observatory of Japan, 2-21-1 Osawa, Mitaka, Tokyo 181-8588, Japan}
\affiliation{Astronomical Science Program, The Graduate University for Advanced Studies, SOKENDAI, 2-21-1 Osawa, Mitaka, Tokyo 181-8588, Japan}
\email{john.livingston@nao.ac.jp}

\author[0000-0001-9626-0613]{Daniel M. Krolikowski}
\affiliation{Steward Observatory, The University of Arizona, 933 N. Cherry Avenue, Tucson, AZ 85721, USA}
\email{krolikowski@arizona.edu}

\author[0000-0001-8511-2981]{Norio Narita}
\affiliation{Komaba Institute for Science, The University of Tokyo, 3-8-1 Komaba, Meguro, Tokyo 153-8902, Japan}
\affiliation{Astrobiology Center, 2-21-1 Osawa, Mitaka, Tokyo 181-8588, Japan}
\affiliation{Instituto de Astrofísica de Canarias (IAC), 38205 La Laguna, Tenerife, Spain}
\email{narita@g.ecc.u-tokyo.ac.jp}

\author[0000-0002-4909-5763]{Akihiko Fukui}
\affiliation{Komaba Institute for Science, The University of Tokyo, 3-8-1 Komaba, Meguro, Tokyo 153-8902, Japan}
\affiliation{Instituto de Astrofísica de Canarias (IAC), 38205 La Laguna, Tenerife, Spain}
\email{afukui@g.ecc.u-tokyo.ac.jp}

\author[0000-0003-3618-7535]{Teruyuki Hirano}
\affiliation{Astrobiology Center, 2-21-1 Osawa, Mitaka, Tokyo 181-8588, Japan}
\affiliation{National Astronomical Observatory of Japan, 2-21-1 Osawa, Mitaka, Tokyo 181-8588, Japan}
\email{teruyuki.hirano@nao.ac.jp}

\author[0000-0001-7246-5438]{Andrew Vanderburg}
\affiliation{Center for Astrophysics \textbar \ Harvard \& Smithsonian, 60 Garden Street, Cambridge, MA 02138, USA}
\affiliation{Department of Physics and Kavli Institute for Astrophysics and Space Research, Massachusetts Institute of Technology, Cambridge, MA 02139, USA}
\email{andrew.m.vanderburg@gmail.com}

\author[0000-0001-9811-568X]{Adam L. Kraus}
\affiliation{Department of Astronomy, The University of Texas at Austin, Austin, TX 78712, USA}
\email{alk@astro.as.utexas.edu} 

\author[0000-0003-2053-0749]{Benjamin M.\ Tofflemire}
\affiliation{SETI Institute, 339 Bernardo Ave., Suite 200, Mountain View, CA 94043, USA}
\affiliation{Department of Astronomy, The University of Texas at Austin, Austin, TX 78712, USA}
\email{tofflemire@utexas.edu}

\author[0000-0001-9158-9276]{Sydney Vach}
\affiliation{\eso}
\email{sydneyvach.astro@gmail.com}

\author[0000-0002-3199-2888]{Sarah Blunt}  
\affiliation{Department of Astronomy, California Institute of Technology, Pasadena, California, USA}
\affiliation{Center for Interdisciplinary Exploration and Research in Astrophysics (CIERA) and Department of Physics and Astronomy, Northwestern University, Evanston, IL 60208, USA} 
\affiliation{Department of Astronomy \& Astrophysics, University of California, Santa Cruz, CA, USA} 
\email{sarah.blunt.3@gmail.com}

\author{Lissa Haskell}  
\affiliation{Department of Physics and Astronomy, The University of North Carolina at Chapel Hill, Chapel Hill, NC 27599, USA} 
\email{lissahaskell@gmail.com}

\begin{abstract}
Despite the wide range of planet-star (mis)alignments in the mature population of transiting exoplanets, the small number of known young transiting planets are nearly all aligned with the rotation axes of their host stars, as determined by the sky-projected obliquity angle. The small number of young systems with measured obliquities limits statistical conclusions. Here we determine the sky-projected obliquity ($\lambda$) of the 3\,Myr transiting planet with a misaligned outer protoplanetary disk, \altplanetname\ (\planetname), using the Rossiter-McLaughlin (RM) effect. Our dataset lacks a pre-transit baseline and ingress, complicating a blind RM fit. Instead, we use contemporaneous spectra and photometry from a mass-measurement campaign to model the stellar activity trend across the transit and provide an external prior on the velocity baseline. We determine $|\lambda|=11.8^{+5.9}_{-5.0}\degree$. Combined with the published rotational velocity of the star, we find a true three-dimensional obliquity of $\psi=15.2^{+7.3}_{-5.7}\degree$. Our result is consistent with an aligned orbit, suggesting the planet remains aligned to its star even though the outer disk is misaligned, though additional RM observations are needed to exclude the low-probability tail of misaligned ($>30$\degree{}) scenarios present in our posterior. 

\end{abstract}


\section{Introduction} \label{sec:intro}


We expect planets to be broadly aligned with the protoplanetary disks from which they formed, as angular momentum is largely conserved during planet formation \citep{Armitage2011}. However, the young (3 Myr) system IRAS 04125+2902 \citep{Barber2024_iras} seemingly contradicts this expectation. It hosts a transiting planet on an edge-on orbit ($i_p=88.3^{+1.2}_{-1.6}\degree$), but a transitional disk that is closer to face on \citep[$i_d\simeq30\degree$;][]{Espaillat2015, Shoshi2025}. To reconcile this, we must assume that, in the narrow window since formation, something warped the outer disk \citep[e.g., a stellar flyby, companion, or late-accretion;][]{Bate2010, Fielding2015, Nealon2025, Huhn2025} or misaligned the planet \citep{Batygin2012, Spalding2014}. 

Both the host star's spin ($i_*>75\degree$) and the wide stellar companion's orbit ($i_c=95^{+11}_{-4}\degree$) are in better agreement with the planet than the disk, suggesting the disk is the misaligned component \citep{Barber2024_iras}. However, these measurements are imprecise, subject to systematics from the stellar properties (challenging for such young systems), and only offer information about the sky-projected inclinations.

Rossiter–McLaughlin \citep[RM;][]{Rossiter1924, McLaughlin1924} 
and Doppler tomography \citep[DT;][]{Marsh1988, CollierCameron2010} 
measurements can fill this gap. These techniques have now been applied to hundreds of systems \citep[e.g.,][]{Albrecht2022, Triaud2018}, revealing a diverse range of spin–orbit angles. Short-period giant planets, which often yield the highest-precision RM measurements, span nearly the full range of possible obliquities, including polar and retrograde configurations \citep{Albrecht2022}.  
Planets in compact multi-planet systems or with smaller radii tend to be more aligned \citep{Albrecht2013, Morton2014, Winn2017, Dawson2018}, and systems around cooler stars are generally aligned, consistent with tidal realignment by convective stellar envelopes \citep{Winn2010, Schlaufman2010, Winn2015}. Large statistical samples of hot Jupiters suggest that misalignment is common for stars hotter than $\sim6250$\,K, while low-mass hosts exhibit a predominance of aligned orbits \citep{Triaud2018}.

Young systems may have some advantages over their older counterparts when measuring the spin-orbit alignments. Young planets tend to be larger \citep[e.g.][]{Mann2016a, Fernandes2022, Vach2024_occ} and their stars have smaller rotation periods (larger rotational broadening), creating a large signal from the planet. While young stellar jitter often complicates planet discovery, for alignment measurements, it is mitigated by a short transit window. RM/DT measurements of six other 10–50 Myr planets \citep[e.g., K2-33, DS Tuc, V1298\,Tau;][]{Hirano2024,Zhou2020,Feinstein2021,Johnson2022} all show low obliquities ($\lambda\lesssim20\degree$), consistent with alignment. This emerging pattern is consistent with young planets forming and remaining aligned within their natal disks and star, either through in situ assembly or via smooth disk migration. However, the sample of young systems with obliquity measurements is too small to generalize; more measurements are needed to make statistical statements about how star-planet alignment varies with time.

In this letter, we expand the sample of young planets with an RM analysis and further explore questions about the young planet IRAS\,04125+2902\,b using radial-velocity observations of the system taken with the \maroonx{} spectrograph and a suite of complementary photometry and spectra. 

\section{Observations}

\subsection{MAROON-X}

\maroonx\ is a high-resolution (R$\sim$85000) spectrograph covering optical wavelengths (500--920nm). The fiber-fed echelle spectrograph splits the light into red (649--920nm) and blue (500--663nm) channels \citep{Seifahrt2018, Seifahrt2020}. We used \maroonx\ to observe a transit of \planetname\ on UT 2024 November 27. Due to the long transit duration, we did not obtain a pre-transit baseline or ingress, but we were able to observe $\sim3.5$ hours of post-transit baseline, totaling 8.65 hours of observations. 

The blue and red channels were reduced separately with a custom pipeline from the \maroonx{} team. The relative radial velocities were extracted using the \texttt{SERVAL} pipeline \citep{Zechmeister2018}. \texttt{SERVAL} creates a stellar template by stacking the spectra and comparing an RV-shifted template to each observation. Long term instrumental drift (on the order of cm~s$^{-1}$~d$^{-1}$) was corrected for using the white light etalon. 

The default RVs provided from \maroonx{} are a weighted mean and standard error across all 56 orders (36 blue and 20 red). We excluded the H$\alpha$ order in both blue and red (due to the strong emission line) and inflated uncertainties using the order-to-order scatter \citep[Equation 15 of][]{Zechmeister2018}. These changes had no impact on the result, but were physically or statistically motivated. We kept blue and red channel RVs separated due to expected differences in limb-darkening and the amplitude of the stellar variability.

Due to the higher airmass at the beginning of night and generally mixed weather conditions, there is higher scatter in the first few RV measurements compared to the later ones. 

\subsection{TESS}

\tess\ observed \starname\ in Sector 86 from 2024 November 21 to 2024 December 18. The target was pre-selected for 20-second, fast cadence observations, with two transits of \planetname\ visible in the dataset. The first transit is simultaneous with the remaining observations. We cut out the 12 hour window surrounding the transit (2460641.58 -- 2460642.08 BJD) to use in the global transit fit (Section \ref{sec:RM}). Because the star is relatively faint, the remaining \tess\ data provides minimal improvement over just fitting the simultaneous data and reduces the risk of contamination from undetected flares or spot crossings in other transits (see Section~\ref{sec:RM}). We extracted the \tess{} light curve and assigned uncertainties to the flux following \cite{Barber2024_iras} and \cite{Vanderburg2019}. The \tess\ data used in this analysis can be found in MAST \citep{tesslc_spoc}.

The \tess{} data is not precise enough to detect effects like spot crossings or small flares at high confidence. There is a perturbation near egress (about 2 hours from mid-transit) that is consistent with a spot crossing, but this is not seen in the more precise ground-based data (and a spot crossing would be stronger in the bluer wavelengths), so we conclude this is random noise. 

\subsection{LCOGT}

We obtained partial observations of the transit (UT 2024 November 27) from three different instruments/telescopes in the LCOGT network; 1) pre-transit to just past mid-transit using the $i'$ filter and QHY600 imager on the 0.4m at McDonald Observatory, 2) the same coverage using the $r'$ filter and Sinistro imager on the 1.0m at McDonald Observatory, and 3) just before egress to post-transit using the MuSCAT3 multi-band imager \citep[simultaneous $g'$, $r'$ $i'$ and $z_s$;][]{Narita2020} on the 2.0m Faulkes Telescope North on Haleakala Observatory. 

Basic reduction for all LCOGT data was done with the \texttt{BANZAI} pipeline \citep{McCully:2018}. For (1) and (2) we built light curves for \starname{} using the \texttt{BANZAI} extracted fluxes and nine comparison stars with similar brightness to \starname{}. We tested for a color term \citep[second-order extinction;][]{Mann2011}, but found it provided no significant improvement. For (3), we performed differential photometry with a customized aperture-photometry pipeline \citep{Fukui2011}. For each band, we re-extracted the fluxes by adopting an optimal aperture radius ($4.8\arcsec$–$6\arcsec$) and selecting the optimal set of comparison stars (2–5 stars).

Weather was mixed at both locations, causing lower-than-expected precision and partial loss of the transit coverage in the photometry. However, the ingress or egress were clearly visible (Figure~\ref{fig:ground}). There was also no evidence of spot crossings or flares during transit that might have complicated the analysis of the velocity data. A small increase in flux is seen in the $g'$ out-of-transit data (4h after mid-transit). While not significant, it is also seen in the \tess{} data (Figure~\ref{fig:ground}) and matches the timing for a small offset in the MAROON-X blue RVs (Figure~\ref{fig:RM}). We attribute this to a small flare. Since this is out of transit, it has no meaningful impact on our result (including or removing that point has no change), but it highlights our ability to detect larger stellar variations that could have impacted the RM result.  

\section{Analysis}

\subsection{Deriving a prior on the velocity trend}\label{sec:hpf}

Young stars are expected to show radial velocity jitter due to spots and other stellar activity. Normally this is fit out using the pre- and post-transit baseline \citep{Montet2020,THYMEIII}, but our \maroonx{} velocities lack any pre-transit baseline. To mitigate this, we derived a prior on the stellar signal over the transit window from a suite of precision RV data taken from the Habitable-zone Planet finder \citep[HPF;][]{Mahadevan2012} and ground-based photometry from LCOGT. These data were taken as part of a campaign to measure the mass of \planetname{}, with additional details in Krolikowski et al. (in prep). To briefly summarize, this includes 92 radial velocities and more than 4700 photometric points across five bands ($g'$, $r'$, $i'$, $Z_s$, and $Y$), both spread over the 2024B observing season (UTC 2024 Oct 24 -- 2025 Feb 19). 

We used the SHO kernel in \texttt{Celerite2} \citep{celerite2} to fit the $i'$-band and the HPF RVs. We optimized the four SHO parameters ($S_0$, $W_0$, $\tau$, and $Q$) on the photometry, then held that kernel fixed while fitting two linear coefficients that convert the prediction into RV space: one to scale the latent GP into RV units ($g$) and one constant offset ($c$) to absorb any zero‐point difference to the measured RVs. The SHO hyperparameters control the shape and coherence of the variability, while the linear coefficients simply map that activity signal into the correct amplitude and baseline in the radial‐velocity domain. This approach assumes the RV and photometric variability have the same underlying cause (mainly surface inhomogeneities). 

We evaluated the trained GP+linear model in a 10-hour window centered at the transit midpoint. This yielded a trend of $-16.9$\,\msd. Using the quasi-periodic kernel \citep{Nicholson2022}, training using $r'$ or $Z_s$ photometric bands, or including the published RVs from \citet{Donati2025} all yielded consistent results on the RV slope ($-5$\,\msd\ to $-20$\,\msd), although errors were much larger when using the RV dataset from \citet{Donati2025}.

Because the HPF data is taken in $zYJ$ band, it will show a shallower slope than we expect to see in the \maroonx{} blue and red channels. 
Assuming the stellar signal is driven by surface inhomogeneities, HPF data will show a weaker slope than \maroonx{}, with the difference driven by the variation in the spot contrast. To account for this, we used the multi-band photometry in a 24-day window centered on the transit (covering two rotation periods). Following \citet{Mori2024}, the difference in amplitude as a function of wavelength provides a strong constrain on difference between the surface and spot temperature. By keep the surface temperature fixed at 4080\,K \citep{Barber2024_iras}, we measured a spot temperature of $3527\pm40$\,K. This is consistent with prior estimates of similar-\teff{} stars \citep[$\Delta T\simeq500$\,K;][]{Herbst2021, Mori2024} and the two-temperature spectral fit in \citet{Barber2024_iras}. 

We estimate linear trends of $-23.8\pm7.3$\,\msd{} (\maroonx{} red channel) and $-31.6\pm9.6$\,\msd{} (\maroonx{} blue channel). The resulting trendline is consistent with ($<1.5\sigma$) the post-transit RVs directly from \maroonx{} (Figure~\ref{fig:RM}). The \maroonx{}-only slope had large uncertainties ($>30$\msd), highlighting the need for the external constraint. The model also predicts only a weak curvature (i.e., 2nd order term) over the transit window, which was untestable with the \maroonx{} data alone because of the narrow baseline, but motivating our decision to model the data with a simple linear trend (Section~\ref{sec:RM}).

\subsection{Global Fit of Photometry and Velocities}\label{sec:RM}

\begin{figure*}
    \centering
    \includegraphics[width=0.98\linewidth]{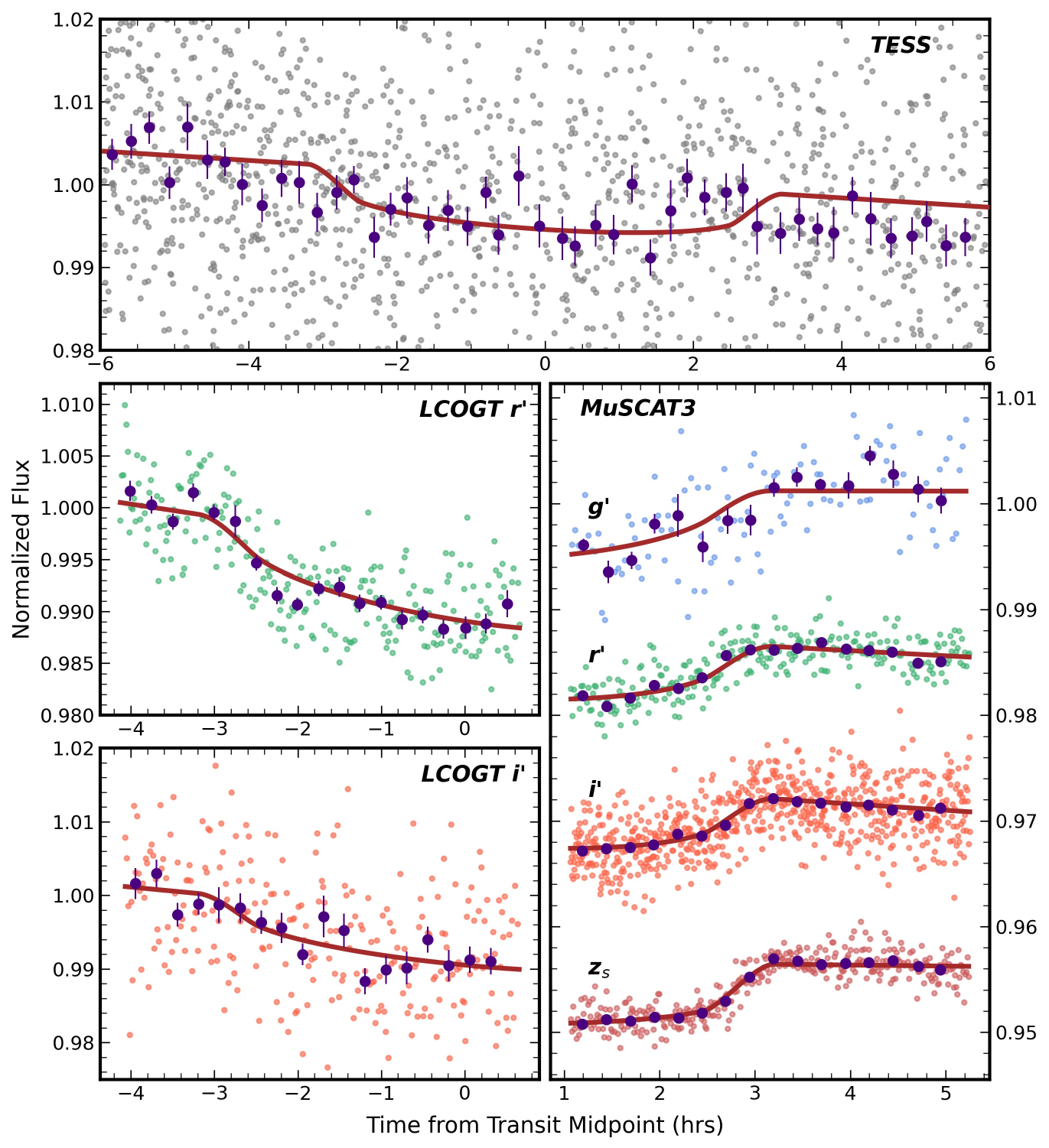}
    \caption{Simultaneous photometric observations of the transit of \planetname\ in \tess\ (top), \lco\ (left), and \muscat\ (right). In all observations, the 15-minute binned data is shown in purple with the raw data shown as the background colored points. The best-fit transit and stellar variability model is shown as the brown line in every filter.}
    \label{fig:ground}
\end{figure*}

The shape of the RM signal is impacted by the planet transit parameters (especially the transit duration). Due to the missed ingress, blindly fitting the RM with transit priors from \cite{Barber2024_iras} caused the model to prefer a too-short transit duration. In order to achieve the most precise parameters, we instead simultaneously fit the \tess\ and \lco\ photometry with the \maroonx\ radial velocities.

Our fit used the model and a significant amount of code from \texttt{MISTTBORN} \citep[MCMC Interface for Synthesis of Transits, Tomography, Binaries, and Others of Relevant Nature;][]{Mann2016a, MISTTBORN}\footnote{\url{https://github.com/captain-exoplanet/misttborn}}. \texttt{MISTTBORN} uses \texttt{emcee} \citep{emcee} to explore parameter space in an MCMC framework and can be used to simultaneously fit transit parameters \citep[using \texttt{BATMAN};][]{BATMAN}, stellar variability, the RM, and an RV trend line. Additional details on the RM implementation in \texttt{MISTTBORN} can be found in \citet{THYMEIII} and \citet{Johnson2022}. 

The implementation of \texttt{MISTTBORN} doesn't allow for simultaneous multi-wavelength photometry and simultaneous multi-wavelength radial velocities, so we wrote an updated wrapper for \texttt{MISTTBORN} adding the following implementations:
\begin{itemize}
    \item Allow for each photometry dataset to be fit with an independent, polynomial trend line.
    \item Update the impact parameter ($b$) prior to allow for constraints on $|b|$ instead of just $b$, and seed starting $b$ values randomly positive or negative.
    \item Allow for multi-wavelength simultaneous radial velocities to be fit with independent RV trend lines and limb-darkening but common RM parameters. 
\end{itemize}

The light curves generally show a downward trend outside of transit (Figure~\ref{fig:ground}), but we split the photometry by instrument and filter to account for differences in telescope systematics and wavelength-dependence on the stellar variability. This resulted in seven total light curves. For each curve, we fit a linear trend line to model the stellar variability in the form
\begin{equation}
    f_{corrected} = f_{raw}-(a + b\times t)
\end{equation}
where $f$ is the normalized flux and $t$ is the number of days since the first data point. The surrounding data in the \tess\ sector and long-term spot monitoring (Section~\ref{sec:hpf}) both indicate a linear term is sufficient over the transit window. Further, with only egress or ingress in the ground-based data, fitting higher order polynomials was impractical. 

We modeled the transit using a \texttt{BATMAN} model \citep{BATMAN}. We fit all light curves with a common transit depth ($R_P/R_*$), transit midpoint ($T_0$), impact parameter ($b$), and stellar density ($\rho_*$). 

We placed priors on $T_0$ and $b$ using the \tess\ transit fit parameters of sectors 19, 43, 44, 59, 70, and 71 presented in \cite{Barber2024_iras} (see Table \ref{tab:THEtable}). Because the target is relatively faint, and the \lco\ data is taken with larger apertures than \tess{}, these priors are relatively weak compared to the constraints provided by the new \lco\ data. In principle, we could fit all data at once without the informed priors, but this would require fitting with a joint GP (to handle long-term variability in \tess{} data) as well as the separate linear terms for the ground-based data. Prior transits also may have undetected spot crossings or flares (especially the longer-cadence data), while we would notice that in the multi-wavelength simultaneous data.

We allowed for positive and negative $b$ values to account for the planet transiting either the northern or southern hemisphere of the star, which is indistinguishable in the photometric transit but affects the shape of the RM. We applied the prior to the absolute value of $b$. Since the transit duration is impacted by the stellar density and orbital eccentricity \citep{Dawson2012,Kipping2014,Van-Eylen2015}, we opted to assume a circular orbit but allowed the stellar density to float with only physical limitations. We initialized the quadratic limb darkening coefficients calculated using \texttt{LDTK} \citep{Parviainen2015_LDTK} and used a Gaussian prior based on differences between different methods or models (0.05--0.1). We locked the planet's period to the value calculated in \cite{Barber2024_iras}.

For each \maroonx\ channel, we fit for two quadratic limb darkening coefficients (calculated using \texttt{LDTK} with a Gaussian prior applied), and a linear trend line in the form
\begin{equation}
    RV_{corrected} = RV_{raw}-(\gamma+\dot\gamma\times t)
\end{equation}
where $t$ is the number of days since the first data point. 

We modeled each channel using a common sky-projected obliquity angle ($\lambda$), rotational broadening (\vsini), and intrinsic width of the Gaussian line profile of individual surface elements ($v_{int}$). The transit parameters (transit midpoint ($T_0$), transit depth ($R_P/R_*$), impact parameter ($b$), stellar density ($\rho_*$)) are fit simultaneously with the photometry. We apply a prior to $\dot\gamma$ using the HPF RV analysis above (Section \ref{sec:hpf}). We put a loose prior on $\gamma$ just to prevent wandering walkers. We initialize and place a prior on $v_{int}$ based on the \maroonx{} instrument profile and an estimate of the macroturbulence from \cite{Brewer_2016}. 

Following the methodology in \cite{Hirano_2011} and \cite{Addison_2013}, we produce a model of the RM using the analytical functions of \vsini\ and $v_{int}$ and taking into account the change in flux due to the planet's transit.

In total the fit included 39 parameters (4 transit parameters, 2 stellar variability parameters per photometry dataset, 2 RV trend parameters per \maroonx\ channel, 3 RM parameters, and 2 limb-darkening parameters per filter/channel). We ran the MCMC fit with 150 walkers for 150,000 steps with a 20\% burn-in. The total run was more than 50 times the autocorrelation time, sufficient for convergence. All parameters and priors are listed in Table~\ref{tab:THEtable}.

\begin{table*}
    \centering
    \caption{Fit parameters and priors}
    \begin{tabular}{lccc}
    \hline
    \hline
    Description & Parameter & Prior$^\alpha$ & Value \\ 
    \hline
    \multicolumn{4}{c}{Transit Parameters}\\
    \hline
    transit mid-point & $T_0$ (BJD-2457000) & $N(3641.8306, 0.005)$ & $3641.8282^{+0.0021}_{-0.0023}$ \\
    planet-to-star radius ratio & $R_P/R_*$ & $U(0,1)$ & $0.0764^{+0.0023}_{-0.0026}$ \\
    impact parameter$^\beta$ & $|b|$ & $N(0.34, 0.24)$ & $0.665^{+0.050}_{-0.068}$ \\
    stellar density & $\rho_*$ ($\rho_\odot$) & $\rho > 0$ & $0.121^{+0.027}_{-0.020}$ \\
    \hline
    \multicolumn{4}{c}{Rossiter-McLaughlin Parameters}\\
    \hline
    rotational broadening & \vsini (\kms) & $N(7.1,0.5)$ & $6.34^{+0.43}_{-0.42}$ \\
    intrinsic width & $v_{int}$ (\kms) & $N(3.3,0.2)$ & $3.35\pm0.20$ \\
    obliquity & $\lambda$ ($\degree$) & $U(-180,180)$ & $11.85^{+5.9}_{-5.0}$ \\
    \hline
    \multicolumn{4}{c}{Photometry Trend Line Parameters}\\
    \hline
    \tess\ flux offset & $a_{\tess}$ (flux) & $\cdots$ & $0.004228^{+0.00066}_{-0.00065}$\\ 
    \tess\ flux slope & $b_{\tess}$ (flux~d$^{-1}$) & $\cdots$ & $-0.0165^{+0.0023}_{-0.0024}$\\
    $\muscat\_{g'}$ flux offset & $a_{\muscat\_{g'}}$ (flux) & $\cdots$ & $0.00081^{+0.00075}_{-0.00075}$\\ 
    $\muscat\_{g'}$ flux slope & $b_{\muscat\_g'}$ (flux~d$^{-1}$) & $\cdots$ & $0.0070^{+0.0072}_{-0.0073}$ \\
    $\muscat\_r'$ flux offset & $a_{\muscat\_r'}$ (flux) & $\cdots$ & $0.00268^{+0.00034}_{-0.00034}$\\ 
    $\muscat\_r'$ flux slope & $b_{\muscat\_r'}$ (flux~d$^{-1}$) & $\cdots$ & $-0.0118^{+0.0028}_{-0.0029}$\\
    $\muscat\_i'$ flux offset & $a_{\muscat\_i'}$ (flux) & $\cdots$ & $0.00348\pm0.00033$\\ 
    $\muscat\_i'$ flux slope & $b_{\muscat\_i'}$ (flux~d$^{-1}$) & $\cdots$ & $-0.0161\pm0.0027$\\
    $\muscat\_z_s$ flux offset & $a_{\muscat\_z_s}$ (flux) & $\cdots$ & $0.00220^{+0.00036}_{-0.00037}$\\ 
    $\muscat\_z_s$ flux slope & $b_{\muscat\_z_s}$ (flux~d$^{-1}$) & $\cdots$ & $-0.0058\pm0.0028$\\
    $\lco\_r'$ flux offset & $a_{\lco\_r'}$ (flux) & $\cdots$ & $0.00053\pm0.00054$\\ 
    $\lco\_r'$ flux slope & $b_{\lco\_r'}$ (flux~d$^{-1}$) & $\cdots$ & $-0.0300\pm0.0048$\\
    $\lco\_i'$ flux offset & $a_{\lco\_i'}$ (flux) & $\cdots$ & $0.0009\pm0.0010$\\ 
    $\lco\_i'$ flux slope & $b_{\lco\_i'}$ (flux~d$^{-1}$) & $\cdots$ & $-0.0227^{+0.0088}_{-0.0089}$\\
    \hline
    \multicolumn{4}{c}{RV Trend Line Parameters}\\
    \hline
    \maroonx{}-blue RV offset & $\gamma_{blue}$ (\ms) & $N(0, 10)$ & $9.3\pm3.5$\\
    \maroonx{}-blue RV slope & $\dot\gamma_{blue}$ (m~s$^{-1}$~d$^{-1}$) & $N(-31.6,9.6)$ & $-33.5^{+9.0}_{-8.9}$ \\
    \maroonx{}-red RV offset & $\gamma_{red}$ (\ms) & $N(0, 10)$ & $3.5^{+2.9}_{-3.0}$\\
    \maroonx{}-red RV slope & $\dot\gamma_{red}$ (m~s$^{-1}$~d$^{-1}$) & $N(-23.8,7.3)$ & $-22.7^{+7.0}_{-6.9}$ \\
    \hline
    \multicolumn{4}{c}{Limb-darkening Coefficients}\\
    \hline
    linear \textit{TESS} & $g_{\tess\_1}$ & $N(0.500,0.1)$ & $0.54\pm0.10$ \\
    quadratic \textit{TESS} & $g_{\tess\_2}$ & $N(0.152,0.05)$ & $0.154^{+0.050}_{-0.049}$ \\
    linear $g'$ & $g_{g'\_1}$ & $N(0.929,0.1)$ & $0.874^{+0.075}_{-0.088}$ \\
    quadratic $g'$ & $g_{g'\_2}$ & $N(0.093,0.05)$ & $0.087^{+0.048}_{-0.045}$ \\
    linear $r'$ & $g_{r'\_1}$ & $N(0.747,0.1)$ & $0.849^{+0.073}_{-0.079}$ \\
    quadratic $r'$ & $g_{r'\_2}$ & $N(0.031,0.05)$ & $0.066^{+0.045}_{-0.039}$ \\
    linear $i'$ & $g_{i'\_1}$ & $N(0.504,0.1)$ & $0.625^{+0.075}_{-0.079}$ \\
    quadratic $i'$ & $g_{i'\_2}$ & $N(0.163,0.05)$ & $0.189\pm0.048$ \\
    linear $Z_s$ & $g_{z_s\_1}$ & $N(0.405,0.1)$ & $0.223^{+0.079}_{-0.081}$ \\
    quadratic $Z_s$ & $g_{z_s\_2}$ & $N(0.194,0.05)$ & $0.159^{+0.048}_{-0.049}$ \\
    linear \maroonx{}-blue & $g_{blue\_1}$ & $N(0.739,0.1)$ & $0.754^{+0.097}_{-0.098}$ \\
    quadratic \maroonx{}-blue & $g_{blue\_2}$ & $N(0.039,0.05)$ & $0.055^{+0.044}_{-0.035}$ \\
    linear \maroonx{}-red & $g_{red\_1}$ & $N(0.489,0.1)$ & $0.55\pm0.10$ \\
    quadratic \maroonx{}-red & $g_{red\_2}$ & $N(0.165,0.05)$ & $0.17\pm0.05$ \\
    \hline
    \multicolumn{4}{l}{$\alpha$ $N(a,b)$ indicates a normal distribution centered at $a$ with a standard deviation of $b$.}\\
    \multicolumn{4}{l}{\hspace*{3.5mm}$U(a,b)$ indicates a uniform distribution from $a$ to $b$.}\\
    \multicolumn{4}{l}{$\beta$ We fit for true $b$ but place the prior on and report $|b|$.}
    \end{tabular}
    \label{tab:THEtable}
\end{table*}

\subsection{Doppler Tomography}
We explored a Doppler tomography (DT) analysis of the \maroonx\ spectra to test whether it could provide an independent constraint on the sky-projected spin–orbit angle. DT directly resolves the moving distortion in the stellar line profile caused by the planet’s shadow \citep{Marsh1988, CollierCameron2010} and is less sensitive to long-term activity or imperfect velocity baselines than traditional Rossiter–McLaughlin (RM) modeling. Given that our dataset lacks a pre-ingress baseline, this technique was particularly appealing.

We attempted a DT analysis using least-squares deconvolution \citep{Donati1997,Kochukhov2010} to extract the time series stellar line profiles. Unfortunately, the combination of modest stellar rotation ($v\sin i_\star \approx 7~\mathrm{km~s^{-1}}$), dense spectral lines (common for K/M stars), and limited per-exposure signal-to-noise ratio means the planet’s Doppler shadow ($\sim0.5~\mathrm{km~s^{-1}}$ wide) was weak and strongly diluted relative to the instrumental and intrinsic line width ($\sim3.3$--$3.5~\mathrm{km~s^{-1}}$). The resulting residual maps showed both random noise and systematics much larger than the expected transit signal and thus did not provide a useful constraint on $\lambda$.

\section{Results}\label{sec:results}

\begin{figure*}
    \centering
    \includegraphics[width=0.98\linewidth]{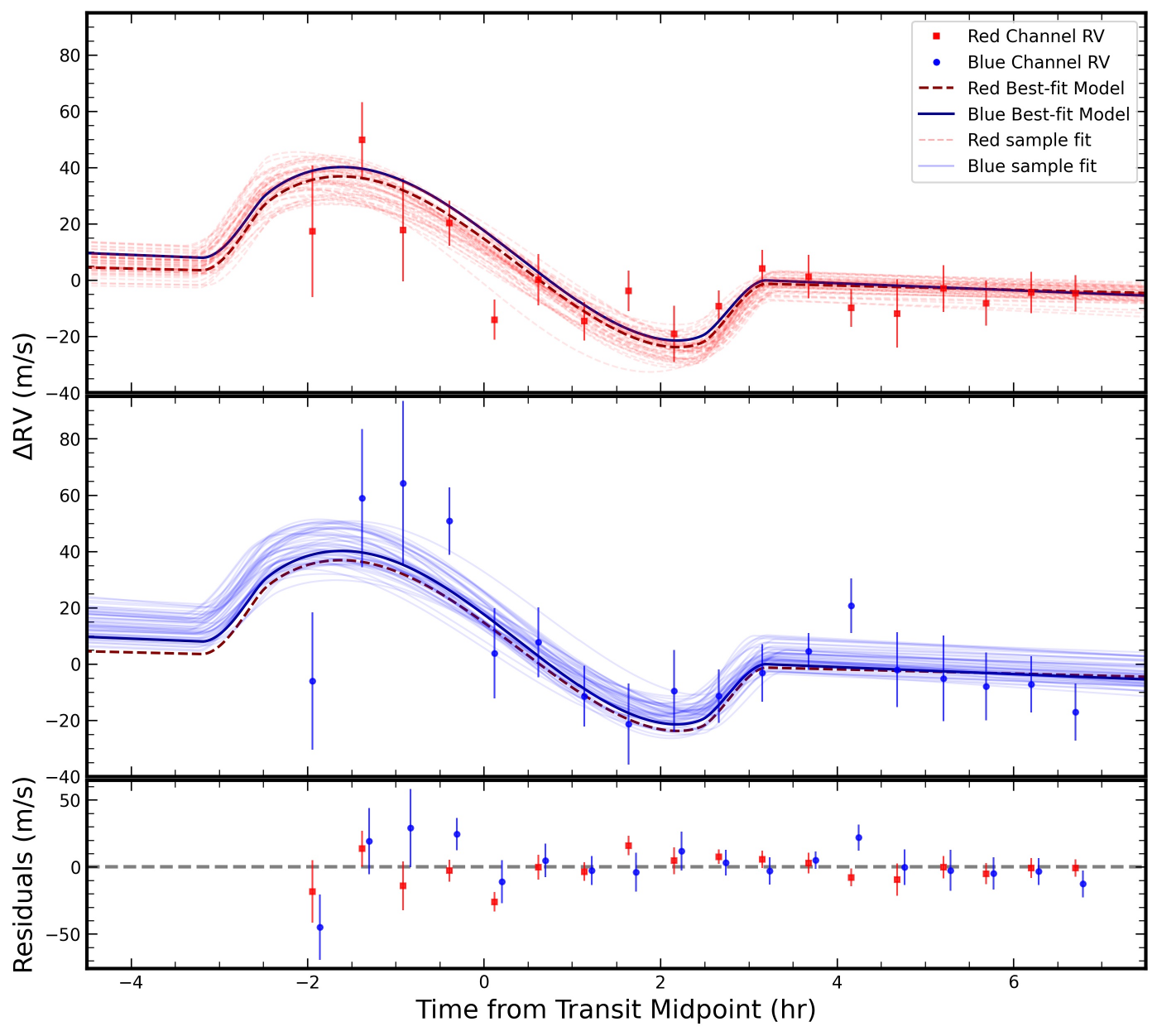}
    \caption{\maroonx\ relative radial velocities from the red (red squares; top) and blue (blue circles; middle) channels. The best-fit RM and stellar RV trend model for each channel is shown as the opaque dotted red and solid blue lines, with 50 random sample fits pulled from the posterior shown as the translucent dotted red and solid blue lines. The bottom plot shows the residuals compared to the best-fit model for each channel, with a 5-minute shift applied to the blue channel RVs for clarity. The red and blue channel fits agree with one another, with the blue RV trend line showing a slightly steeper slope, as expected.}
    \label{fig:RM}
\end{figure*}

We present the best-fit parameters in Table \ref{tab:THEtable} and show the resulting transit fits in Figure \ref{fig:ground} and RM fit in Figure \ref{fig:RM}. The resulting transit parameters are consistent with those found in \citep{Barber2024_iras}, although generally more precise. \cite{Barber2024_iras} reports \tess-only transit fit parameters, while here we simultaneously fit multi-wavelength photometry from higher precision instruments. This increased the overall precision of the $\rho_*$ and $b$ parameters (which agree to discovery parameters to 1.5$\sigma$ and 1.3$\sigma$, respectively). $R_P/R_*$ has similar precision to the discovery value and agrees to 2.2$\sigma$. Some deviation from the discovery paper may be due to differences in our analysis, changes in the spot pattern over time, and dilution from the $\Delta T=2.8$\,mag companion (unresolved in \tess{} but resolved in ground-based data).

We found a final obliquity of $|\lambda|=11.8^{+5.9}_{-5.0}$\degree, which is consistent with alignment or a small misalignment ($<20^\circ$) between the stellar spin axis and the planetary orbit. We computed the true three-dimensional obliquity ($\psi$) from the standard relation:
\begin{equation}
\cos \psi = \cos i_\star \, \cos i_p + \sin i_\star \, \sin i_p \, \cos \lambda.
\end{equation}
Using $i_p$ and $\lambda$ estimated from our transit-fits above, and the posterior on $i_*$ from \citet{Barber2024_iras}, we derive $\psi=15.2^{+7.3}_{-5.7}\degree$.

\begin{figure*}
    \centering
    \includegraphics[width=0.99\linewidth]{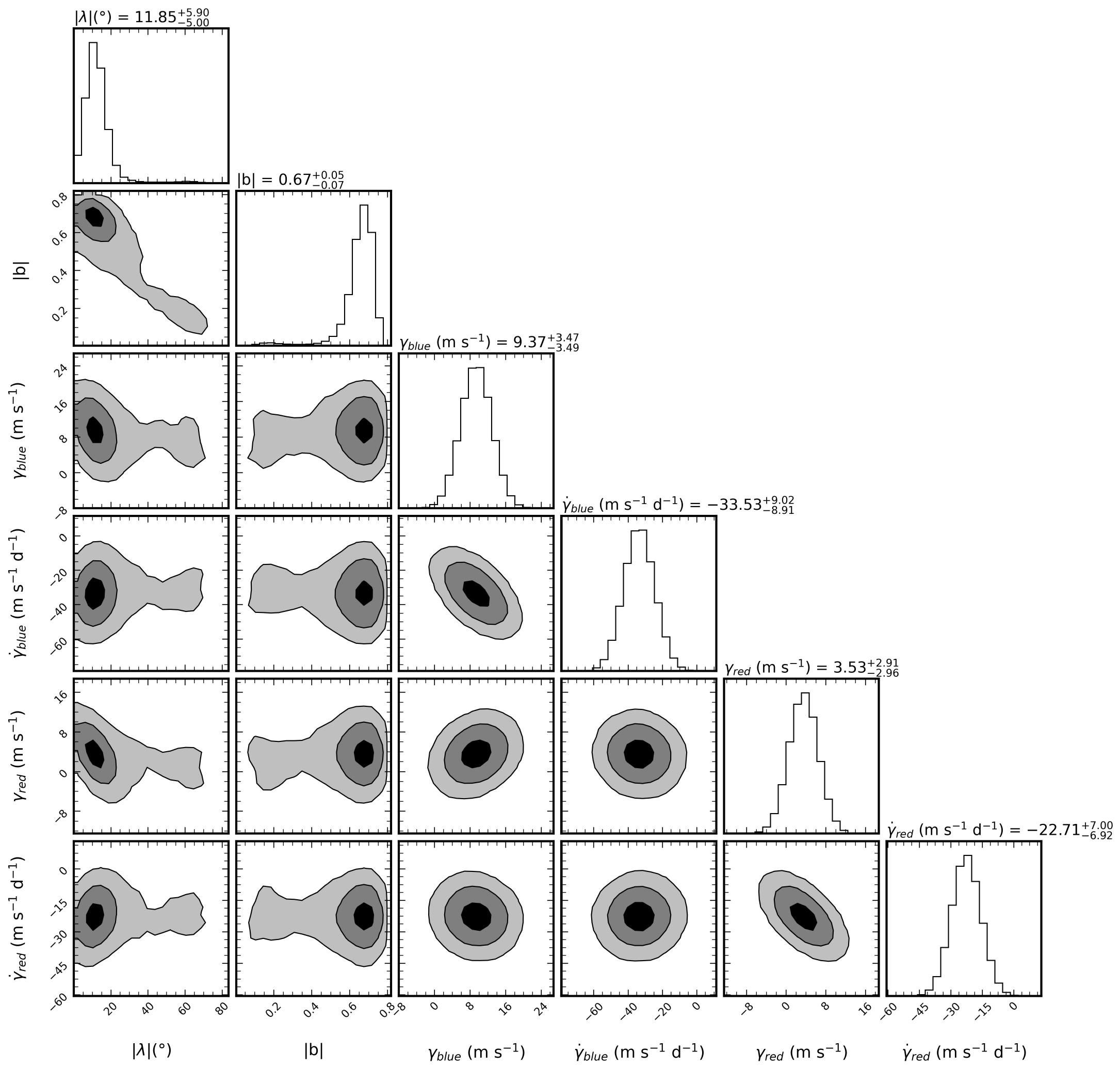}
    \caption{Corner plot of a subset of the RM parameters. The contours show the 1, 2, and 3$\sigma$ levels. Only 99\% of the distribution is shown for clarity. The sky-projected obliquity angle ($\lambda$) is correlated with the impact parameter ($b$) but seems to be minimally impacted by the the RV trend line parameters ($\gamma$ and $\dot\gamma$), with the exemption of the largest $\lambda$ values (with a near-flat RM signal near egress) corresponding to a tighter RV trend line solution. Remaining, unshown parameters are similarly gaussian distributed with little correlation between parameters.}
    \label{fig:corner}
\end{figure*}

Due to the effects on the resulting transit duration, the impact parameter seems to be most correlated with $\lambda$ (Figure~\ref{fig:corner}), with small values of $b$ ($b\lesssim0.4$) straying to the higher end of the $\lambda$ distribution ($|\lambda|\gtrsim40\degree$).

Relaxing the prior on $\dot\gamma$ (tripling the width) has a small effect on the final $\lambda$ value. The best-fit $\lambda$ agrees with the above fit to 1$\sigma$ ($|\lambda|=14.4^{+6.2}_{-5.5}\degree$). 

We tested for signs that our result was impacted by activity from the host star. The rotation period of the star is long (11.3\,days) compared to the window here ($\simeq8$ hours), but flares and other short-lived events can matter on this timescale. Such stellar signals are chromatic, generally stronger in the blue and in orders with strong emission features. To test for this, we computed $\chi^2/N$ between the best-fit model above and the individual RVs from each of the 54 orders used in our analysis. There is no trend with wavelength, and orders with emission lines (e.g., H$\alpha$, calcium triplet) have $\chi^2/N$ consistent with the overall distribution. 

We similarly find no wavelength dependency trends using the simultaneous photometry. Stellar variations should show up in the bluer ($r'$ and $g'$) photometry. Although $g'$ data shows higher uncertainties, the fit is reasonable, with the only odd point at 4-hours from mid-transit. This matches an usually high RV from \maroonx{}-blue, indicating a small stellar flare. However, this is outside the transit, and removing this point had no impact on the result.

\section{Summary and Discussion}
Here we determine the sky-projected spin-orbit angle of \altplanetname\ using radial velocities from \maroonx. Although the observations missed ingress, we use the suite of external data, including multi-band photometry of the transit and velocities from the 2024B HPF radial velocity campaign to place priors on the velocity trend and confirm that a linear model is sufficient to explain the out-of-transit variations. We determine a sky-projected obliquity of $|\lambda|=11.8^{+5.9}_{-5.0}\degree$ and a true three-dimensional obliquity of $\psi=15.2^{+7.3}_{-5.7}\degree$. Loosening our input priors yields a consistent $|\lambda|=14.4^{+6.2}_{-5.5}\degree$, corresponding to $\psi=17.3^{+6.9}_{-6.0}\degree$. This is consistent with an aligned orbit or a small misalignment. 

The youth of the star raises the possibility of stellar activity weakening or enhancing the signal. We find no evidence of this, other than the spot-induced trend-line. No chromaticity is seen in the RM signal, and the photometry does not show any evidence of stellar variations other than the rotation-driven trendline (which is included in our fit).  

One important caveat to our results is the long tail on the $\lambda$ posterior (Figure~\ref{fig:corner}), which reaches to $\simeq60\degree$. This is a small part of the probability distribution, but it does indicate that we cannot rule out a highly misaligned system at $>3\sigma$. The tail corresponds to low values of $b$, which are disfavored (but not totally ruled out) by the transit shape. This is complicated by stellar variability, coverage of the \maroonx{} data (missing ingress), and the larger uncertainties in the first 2-4 epochs (taken at higher airmass). Additional RM observations and high-precision transit photometry of \planetname{} would be invaluable to confirm the result and provide a tighter constraint on $b$ (and hence $\lambda$). 

Had the planet shown a significant misalignment (high $|\lambda|$), the inclination would still be off from the outer disk. This would suggest a massive disruption event that impacted the planet and disk in different ways. Alignment, however, favors interactions that {\it only} impact the disk while the planet remains tied to the rotational direction of the star. Late accretion, for example, would predict this exact outcome \citep{Huhn2025}. There's some evidence that such outer-disk misalignments are common \citep{Ansdell2020, Fields2025}, which favors further searches for transiting planets in systems with transitional disks, which would in turn provide more information about their origin.

\begin{figure*}
    \centering
    \includegraphics[width=0.99\linewidth]{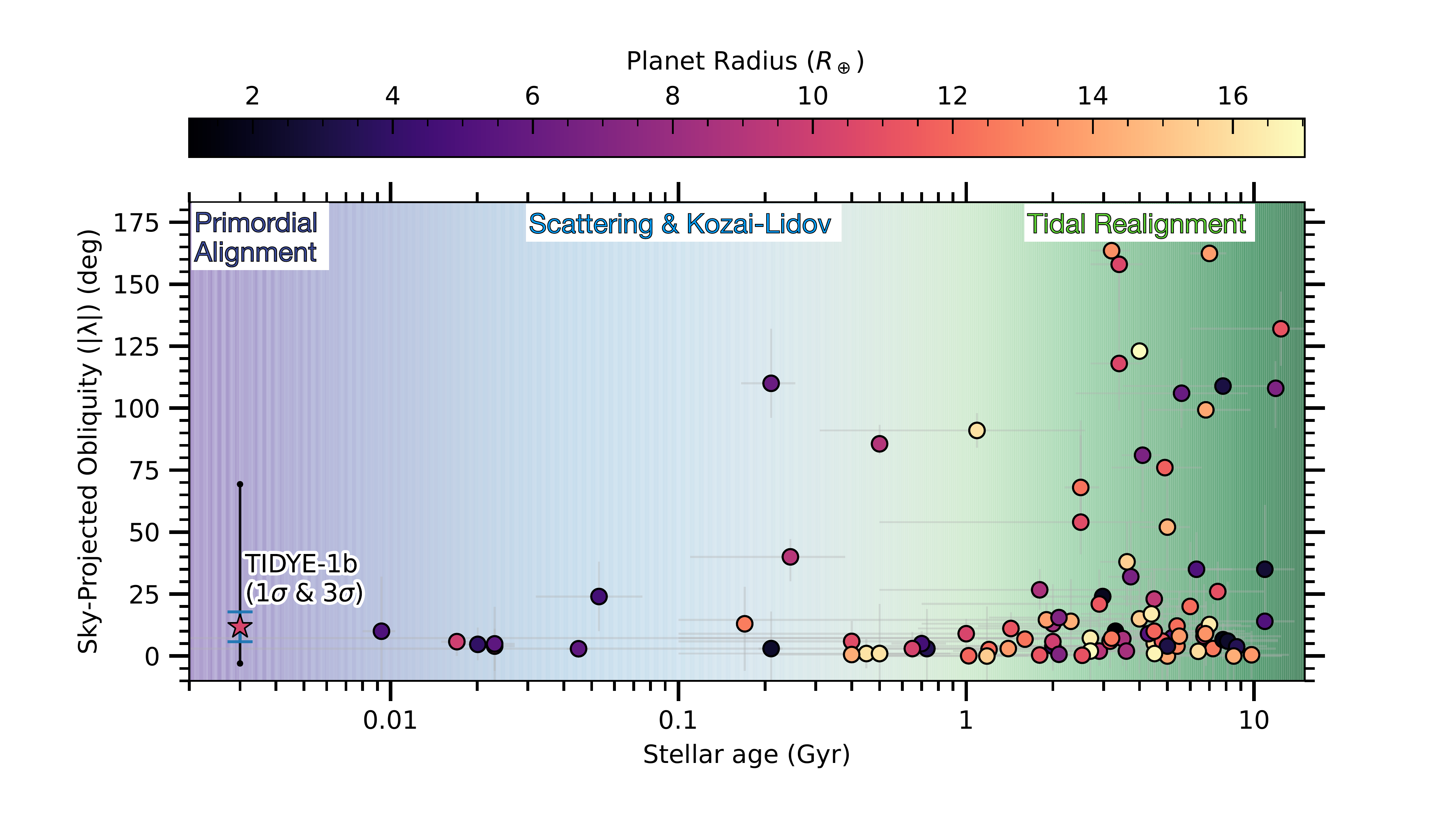}
    \caption{Distribution of sky-projected obliquity angles ($|\lambda|$) for transiting planets across system ages \citep{southworth2011}\footnote{\url{https://www.astro.keele.ac.uk/jkt/tepcat/obliquity.html}}. This only considers planets with masses below 10$M_J$ (or no measured mass), \teff\,$<6250$\,K, $\sigma_\lambda<25^\circ$, and those with literature ages \citep[taken from the NASA Exoplanet archive;][]{exoplanetArchive}. All points are colored by the planet radius. Young systems tend to be aligned but the difference is not significant. \altplanetname\ (\planetname; star) is consistent with the few other $<0.1$\,Gyr systems. The background shading represents rough approximations of the timescales for processes of interest for the SOYSAUCE survey.}
    \label{fig:sample}
\end{figure*}

\begin{table*}[]
    \centering
    \caption{Sky-projected obliquity angles for transiting planets $<$300\,Myr.}
    \begin{tabular}{lccccc}
    \hline
    \hline
    Planet Name & Age & Age Ref. & Sky-Projected Obliquity & 3-D Obliquity & Obliquity Ref. \\
                & (Myr) &        & ($\lambda$, $\degree$) &    ($\psi$, $\degree$)           & \\
    \hline
    TIDYE-1\,b & $3.3^{+0.6}_{-0.5}$ & \cite{Barber2024_iras} & $11.8^{+5.9}_{-5.0}$ & $15.2^{+7.3}_{-5.7}$ & This work \\
    K2-33\,b      & $9.3^{+1.1}_{-1.3}$ & \cite{Mann2016_k233} & $-10^{+22}_{-24}$ & $\cdots$ & \cite{Hirano2024} \\
    HIP 67522\,b & $17\pm2$ & \cite{THYMEII} & $-5.8^{+2.8}_{-5.7}$ & ${20.2}^{+10.3}_{-8.7}$ & \cite{Heitzmann2021} \\
    AU Mic\,b & $20.1^{+2.5}_{-2.4}$ & \cite{Wittrock2023} & $-4.7^{+6.8}_{-6.4}$ & $\cdots$ & \cite{Hirano2020} \\
    AU Mic\,c & $20.1^{+2.5}_{-2.4}$ & \cite{Wittrock2023} & $67.8^{+31.7}_{-49.0}$ & $\cdots$ & \cite{Yu2025} \\
    V1298 Tau\,b & $23\pm4$ & \cite{David2019b,David2019a} & $4^{+7}_{-10}$ & $8^{+ 4}_{-7}$ & \cite{Johnson2022} \\
    V1298 Tau\,c & $23\pm4$ & \cite{David2019b,David2019a} & $4.9^{+15.0}_{-15.1}$ & $\cdots$ & \cite{Feinstein2021} \\
    DS Tuc A\,b & $45\pm4$ & \cite{Newton2019_DSTuc} & $2.93^{+0.88}_{-0.87}$ & $\cdots$ & \cite{Zhou2020} \\
    TOI 942\,b & $53^{+22}_{-21}$ & \cite{Wirth2021} & $1^{+41}_{-33}$ & $2^{+27}_{-33}$ & \cite{Wirth2021} \\
    TOI 942\,c & $53^{+22}_{-21}$ & \cite{Wirth2021} & $24\pm14$ & $<43$ & \cite{Teng2024} \\ 
    Qatar-4\,b & $170\pm10$ & \cite{Alsubai2017_qatar345} & $-13^{+15}_{-19}$ & $32^{+14}_{-13}$ & \cite{Zak2025} \\
    Kepler-63\,b & $210\pm45$ & \cite{SanchisOjeda2013} & $-110^{+22}_{-14}$ & $104^{+9}_{-14}$ & \cite{SanchisOjeda2013} \\
    TOI-2076\,b & $210\pm20$ & \cite{Barber2025_toi2076e} & $-3^{+15}_{-16}$ & $18^{+10}_{-9}$ & \cite{Frazier2023} \\
    TOI-1268\,b & $245\pm135$ & \cite{Subjak2022} & $40.0^{+7.2}_{-9.9} $ & $\cdots$ & \cite{Dong2022}\\ 
    \hline
    \end{tabular}
    \label{tab:currentSample}
\end{table*}

\subsection{The SOYSAUCE survey}

Simple angular-momentum arguments suggest that most planets should form and remain aligned with their host star's spin ($|\lambda|\lesssim20^\circ$). A variety of formative and evolutionary processes can generate misaligned systems (Figure~\ref{fig:sample}). Theoretical work shows that primordial disk warping and early magnetic or turbulent interactions can imprint spin–orbit misalignments even before gas dispersal \citep[$\lesssim$10\,Myr;][]{Bate2010,Lai2011, Batygin2012, Kley2012,Spalding2015, Fielding2015}. From 10–200\,Myr, after the disk dissipates, secular interactions dominate \citep[e.g. planet-planet scattering and secular chaos;][]{Weidenschilling1996,Rasio1996,Wu2011,Naoz2011,Naoz2013}, and Kozai-Lidov \citep{Kozai1962,Lidov1962} cycles may tilt the planet's orbit \citep[e.g.,][]{Wu2003, Fabrycky2007, Nagasawa2008, Lithwick2011,Petrovich2015,Naoz2016}. Beyond a few hundred Myr, tides inside the stellar convective envelope may dampen misalignments \citep[tidal re-alignment; e.g.,][]{Albrecht2012, Winn2015,Dawson2018}. Timescales for these mechanisms are impacted by the system's parameters (e.g. stellar mass and radius and planetary mass) and the presence of multiple planets or stars in the system.

The youngest ($\lesssim$50\,Myr) planets observed so far, including \planetname, show low sky-projected obliquities (Figure~\ref{fig:sample}, Table~\ref{tab:currentSample}). This pattern is consistent with disk-driven migration preserving alignment and hints that aligned configurations are more common at younger ages. A preliminary two-sample comparison using Kolmogorov–Smirnov \citep{Kolmogorov1933,Smirnov1936,Smirnov1939} and Anderson-Darling \citep{AndersonDarling1954} tests (KS/AD tests) across the age bins above does not yet yield a significant difference; we are limited by small-number statistics and population differences, as young planets are typically larger (“puffier”) than mature planets \citep[e.g.,][]{Mann2016a,Fernandes2022,Vach2024_occ} and relatively few have secure masses \citep[e.g.,][]{Thao2024_featherweight,Barat2025}, complicating fair comparisons across size/mass regimes.

We performed a simplified simulation to estimate how many additional systems are required. Assuming that planets are born aligned and remain so through $\sim\!100$\,Myr, and treating the current old sample as the reference misalignment rate, we would need at least 10 additional $<$100\,Myr systems with $\sigma_\lambda\!\le\!25^\circ$ to achieve a \textbf{3\boldmath$\sigma$} separation between the young and old misalignment fractions. We only considered literature values with $\sigma_{\lambda}<20^\circ$, as low-precision measurements like those for AU Mic c \citep[$\lambda = 68^{+32}_{-49}\degree$;][]{Yu2025} provide ambiguous conclusions. We also tested a more complicated situation where planets $<$20\,Myr are all aligned, and 20\% of them become misaligned from 20-300\,Myr, and about half of those becoming re-aligned by 2\,Gyr. Separating this model from a flat misalignment fraction at all ages would require $\lambda$ measurements for an additional 19 $<$300\,Myr systems. This regime ($<300$\,Myr) also corresponds to the range (in log-age) with relatively few measurements overall (Figure~\ref{fig:sample} and Table~\ref{tab:currentSample}).

The SOYSAUCE survey was designed with this in mind. Our goals are twofold: 1) measure sky-projected obliquity angles ($\sigma_\lambda \lesssim25\degree$) for enough $\lesssim$500\,Myr planets to test models of dynamical evolution and 2) increase the number of young planets with preliminary atmospheric characterizations from the same dataset. For the former, the science goal is to test dynamical evolution similar to our simulations above. At least five additional $<$300\,Myr systems already have RM/DT observations obtained or scheduled and are pending publication. Combined with ongoing searches for young transiting planets \citep[e.g.,][]{Newton2019_DSTuc, Fernandes2022, Vach2024_tic4343, Barber2024_hipc} and improvements in age-dating field stars \citep[e.g.,][]{2023ApJ...947L...3B, BarberMann2023, Jeffries2023}, the sample required for (1) is within reach in the next few years.

The second goal is to search for strong atomic features in the transmission spectra derived from the same RM/DT data, taking advantage of strong feature strength seen in some young planets \citep{Thao2024_featherweight, Barat2025}. In the best cases, this can provide classification-level constraints on the mass \citep[e.g. small planet progenitor vs true gas giant;][]{deWit2025} that can help with the dynamics program. At minimum, these can provide some of the best targets for further follow-up with {\it JWST} or ARIEL in terms of feature strength.

\section*{Acknowledgments}
The authors would like to thank Halee and Bandit for their continual, invaluable, scientific input. M.G.B. was supported by the NSF Graduate Research Fellowship (DGE-2040435) and a grant from NASA's Exoplanet research program (80NSSC25K7148). A.W.M. was supported by grants from the NSF's CAREER program (AST-2143763) and TESS Guest Investigator program (80NSSC25K7903).

This work is partly supported by JSPS KAKENHI Grant Numbers JP24H00017, 
JP24K17083, 
JP24K00689, 
JP25K17450, 
JSPS Grant-in-Aid for JSPS Fellows Grant Number JP24KJ0241, 
and JSPS Bilateral Program Number JPJSBP120249910.

This paper makes use of data collected by the TESS mission. Funding for the \tess\ mission is provided by NASA’s Science Mission Directorate. We acknowledge the use of public \tess\ data from pipelines at the \tess\ Science Office and at the \tess\ Science Processing Operations Center. Resources supporting this work were provided by the NASA High-End Computing (HEC) Program through the NASA Advanced Supercomputing (NAS) Division at Ames Research Center for the production of the SPOC data products. \tess\ data presented in this paper were obtained from the Mikulski Archive for Space Telescopes (MAST) at the Space Telescope Science Institute.

This work makes use of observations from the LCOGT network. Part of the LCOGT telescope time was granted by NOIRLab through the Mid-Scale Innovations Program (MSIP). MSIP is funded by NSF. Some data in the paper are based on observations made with the MuSCAT3 instrument, developed by the Astrobiology Center (ABC) in Japan, the University of Tokyo, and Las Cumbres Observatory (LCOGT). MuSCAT3 was developed with financial support by JSPS KAKENHI (JP18H05439) and JST PRESTO (JPMJPR1775), and is located at the Faulkes Telescope North on Maui, HI (USA), operated by LCOGT.

This work makes use of observations obtained with the Hobby–Eberly Telescope (HET). We are very grateful for the resident astronomers and telescope operators at the HET for help in planning HPF observations. The HET is a joint project of the University of Texas at Austin, the Pennsylvania State University, Ludwig-Maximilians-Universität München, and Georg-August Universität Gottingen. The HET is named in honor of its principal benefactors, William P. Hobby and Robert E. Eberly.

This work makes use of observations obtained with the Habitable-zone Planet Finder Spectrograph. We thank the HPF team for providing the high-quality data products necessary for this work.

\bibliographystyle{aasjournal}
\bibliography{planetSearch}

\end{document}